# Programming Nonlinear Propagation for Efficient Optical Learning Machines

Ilker Oguz[1,2], Jih-Liang Hsieh[1,2], Niyazi Ulas Dinc[1,2], Uğur Teğin[1,2], Mustafa Yildirim[1], Carlo Gigli[2], Christophe Moser[1] and Demetri Psaltis[2]

[1] Laboratory of Applied Photonics Devices, Ecole Polytechnique Fédérale de Lausanne (EPFL), Switzerland

[2] Optics Laboratory, Ecole Polytechnique Fédérale de Lausanne (EPFL), Switzerland

## Abstract

The ever-increasing demand for processing data with larger machine learning models requires more efficient hardware solutions due to limitations such as power dissipation and scalability. Optics is a promising contender for providing lower power computation since light propagation through a non-absorbing medium is a lossless operation. However, to carry out useful and efficient computations with light, generating and controlling nonlinearity optically is a necessity that is still elusive. Multimode fibers (MMF) have been shown that they can provide nonlinear effects with microwatts of average power while maintaining parallelism and low loss.

In this work, we propose an optical neural network architecture, which performs nonlinear optical computation by controlling the propagation of ultrashort pulses in MMF by wavefront shaping. With a surrogate model, optimal sets of parameters are found to program this optical computer for different tasks with minimal utilization of an electronic computer. We show a remarkable decrease of 97% in the number of model parameters, which leads to an overall 99% digital operation reduction compared to an equivalently performing digital neural network. We further demonstrate that a fully optical implementation can also be performed with competitive accuracies.

## Introduction

Machine learning architectures have come to be dominated by artificial neural networks (ANN). There are several reasons why this architecture is used so broadly. Initially, their similarity with biological neural networks[1] has provided strong motivation to explore ANNs. At the same time, the fact that ANNs are universal machines[2], which are able to approximate any function, breeds confidence that ANNs can carry out useful and difficult tasks. Perhaps most significantly, the fact that error back propagation[3] has proven very effective in training such networks catapulted their application to a wide variety of problems. Ever-larger networks[4] have been adopted for tackling challenging tasks[5]. Empirically it has been found that larger networks tend to perform better given a sufficiently large database of training examples. This has led to a "bigger is better" mentality[6] even though theoretically we expect that networks with as few degrees of freedom as possible to generalize better[7]. However, the disadvantage of this mentality is the energy required to train and use very large networks. For instance, only training the language model GPT-3, which has 175 billion parameters, consumed 1.3 GWh of electricity which is the energy required to fully charge 13'000 Tesla Model S cars[8]. Optics can help overcome this downside since light propagation through a non-absorbing, non-scattering medium is a lossless linear operation.

Several approaches have been reported for the optical realization of ANNs. Wavefront shaping by diffractive surfaces or modulators, followed by propagation can implement ANNs and perform different tasks such as classification and imaging [9–13]. Silicon photonics technology also allows the realizations of reconfigurable optical computing with much smaller dimensions. Control structures such as Mach-Zehnder modulators[14] and micro-ring resonators[15] can precisely manipulate light inside waveguides as in building blocks of ANNs, performing operations such as multiplication or addition. However, their constraint to two dimensions diminishes the intrinsic three-dimensional scalability of optics[16]. Additionally, in both of those domains, nonlinearity is obtained generally by optoelectronic devices, which impairs energy efficiency and speed.

Contrary to the update of individual weights of ANNs, the reservoir computing approach requires a very large number of neurons with random and fixed weights without their precise control, combined with only a single layer of reconfigurable readout weights (RW) trained to map the output of this complex transform to a desired output[17]. This paradigm of computing fits well with optics and was implemented with different optical devices[18–32]. The necessary nonlinearity between neurons could be introduced by electronic feedback[18], optoelectronic conversions[27], saturable absorption[30], and second-harmonic generation[33]. Moreover, the Kerr effect inside both single-mode [25] and multi-mode[34] fibers was shown recently to be an effective reservoir for computing systems. Another set of studies showed also that the nonlinear interactions inside MMFs are tunable by wavefront shaping[35–37].

In this paper, we present an optical ANN architecture, which combines a relatively small number of digitally implemented parameters to control the very complex spatiotemporal transformation realized in an MMF. Our experimental studies show that with a spatial light modulator (SLM) by simultaneously shaping light with data and a fixed programming pattern, we can induce the non-linear transformation inside the MMF to perform desired computations. We find that the optimization of a small number of programming parameters (PP), around 50 in our experiments, results in a remarkable performance of the optical computer. For instance, we will show that the system with approximately 2000 total parameters (TP), PPs and RWs combined, performs as well as a digital ANN with over 400,000 parameters for the face classification task[38].

Moreover, we demonstrate that the same method can be used to program the propagation inside MMF to perform all-optical classification without the digital readout stage. In this case, the classification can be directly read out with a simple beam location sensor, further decreasing the number of TPs.

# Results

## Programming Fiber Transform for Higher Classification Performance

Our method consists of a nonlinear optical transformation and a programming algorithm. The transform is the propagation of spatially modulated laser pulses through a graded-index MMF. The nonlinear propagation of an ultrashort pulse inside an MMF is a highly complex process that entails spatial and temporal interactions of electromagnetic waves coupled to hundreds of different propagation modes (see Supplementary Note 1), such that modelling the transform of a single pulse on the setup provided in Figure1 would take 50 minutes with a graphics processing unit (GPU)[39]. The physical optical system carries out this complex spatiotemporal transformation "effortlessly". The transformation is programmed with a relatively small number of PPs. PPs are selected to customize the MMF processor for the specific task by wavefront shaping, controlling the optical power, and the placement of the data and diffraction angle on the SLM. The combination of wavefront shaping and data encoding as shown in Figure 1 can be formalized as follows, where the complex encoded value on the SLM $E_{SLM}^k(x,y)$ is given by:

$$E_{SLM}^k(x,y) = WF(x,y) * \exp\left(iD^k(x,y)\right) = |WF(x,y)| * \exp(i(D^k(x,y) + Arg(WF(x,y)))) \quad \text{Eqn.1}$$

The amplitude modulation with the phase-only SLM is realized by modifying the strength of a blazed grating (see Methods), whereas the data ($D^k(x,y)$) is encoded as a phase pattern. The wavefront controlling shape ($WF(x,y)$) is a complex combination of N different linearly polarized fiber modes, $F_n(x,y)$, and the coefficient of each mode is controlled by two parameters, for real and imaginary parts. Therefore, the portion of the wavefront shape that contains the 2N PPs ($a_n$) is

$$WF(x,y) = \sum_{n=1}^{N}(a_{2n-1} + i * a_{2n}) * F_n(x,y). \quad \text{Eqn.2}$$

During the programming, the system processes the data for different selections of PPs, determines the output stage (RWs) and a surrogate optimization algorithm finds the best performing set of PPs. This optimization is similar to the hyperparameter optimization process in ANNs.

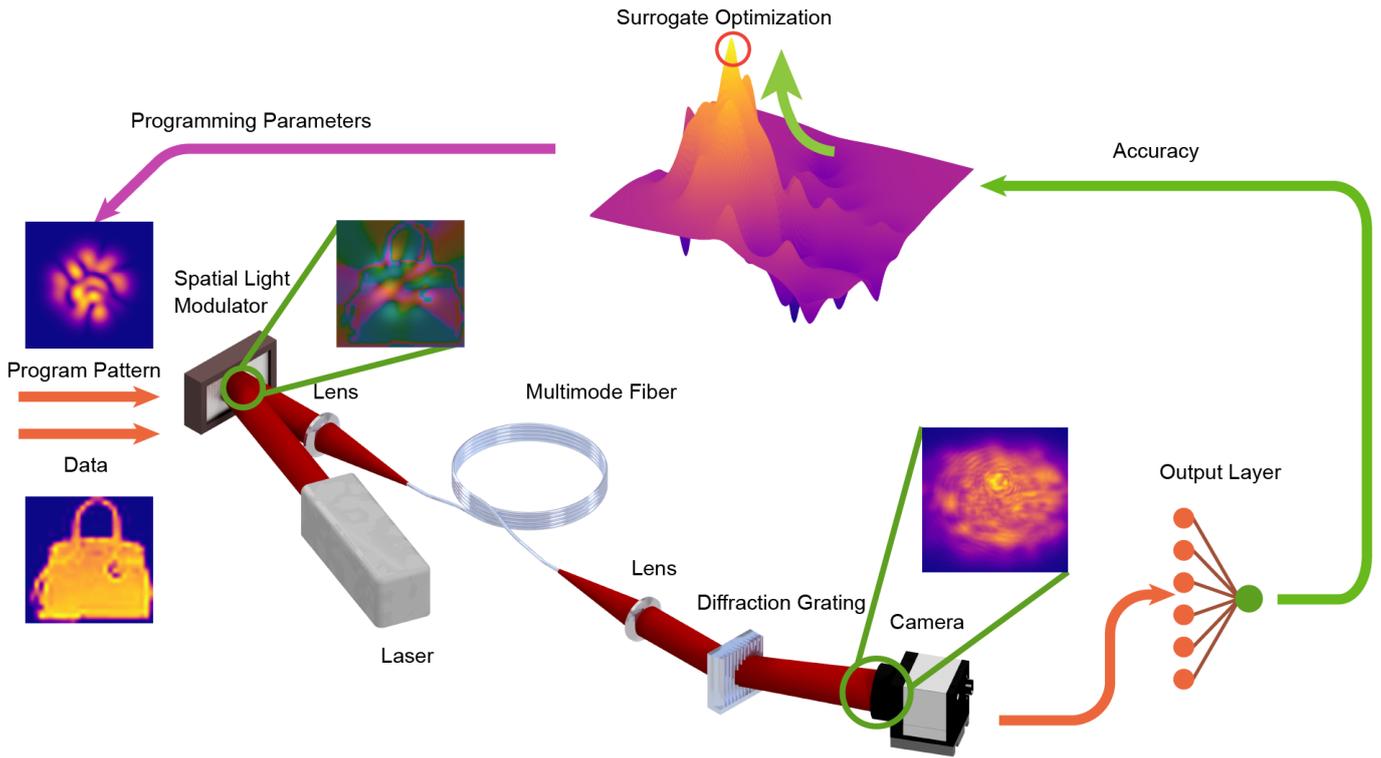

*Figure 1 The experiment flow for programming optical propagation for a computational task*

The peak intensity of the pulsed laser beam is treated as one of the PPs, whose experimental effect is analyzed in Supplementary Note 3, and is optimized through the surrogate model. The laser beam is sent onto a reflective SLM which encodes the data to be transformed by the optical system and combines them with a complex pattern containing the PPs as described in Equation 1. Even though other bases for controlling the wavefront can also be effective, we used the fact that the fiber, $F_n(x, y)$, since they are interpretable, orthonormal, and guaranteed to be within the fiber's numerical aperture.

Once the PPs are determined, the data portion of the modulation is changing with every sample, while the programming part stays the same. The nonlinear nature of the optical transform implies that the PPs modify the nonlinear transform that the dataset undergoes. Upon exiting the fiber, the beam is collimated with a lens and sent to a blazed grating. The dispersion due to the grating leads to a camera recording in which both spatial and temporal characteristics of the output beam are present.

The classification accuracies are calculated by training a simple regularized linear regression algorithm on the pixel intensity values of the recorded images on the camera in Figure 1. The linear regression maps the recorded image to classification results by pointwise multiplication with RWs and summation. Finally, the pairs of PPs and corresponding task performance of the system are supplied to the surrogate algorithm that optimizes the optical transform. After acquiring the performance metric for different sets of PPs, the surrogate optimization algorithm creates mappings between the performance of the system and any given set of PPs. The surrogate algorithm continuously refines the model and increases the performance on the task at the same time. Fig 2. illustrates such an experiment, where the data transform with the MMF was programmed for a higher classification accuracy on a small subset (2%) of the Fashion-MNIST Dataset, which consists of 1200 training images and 300 test images of 10 different classes of fashion items[40].

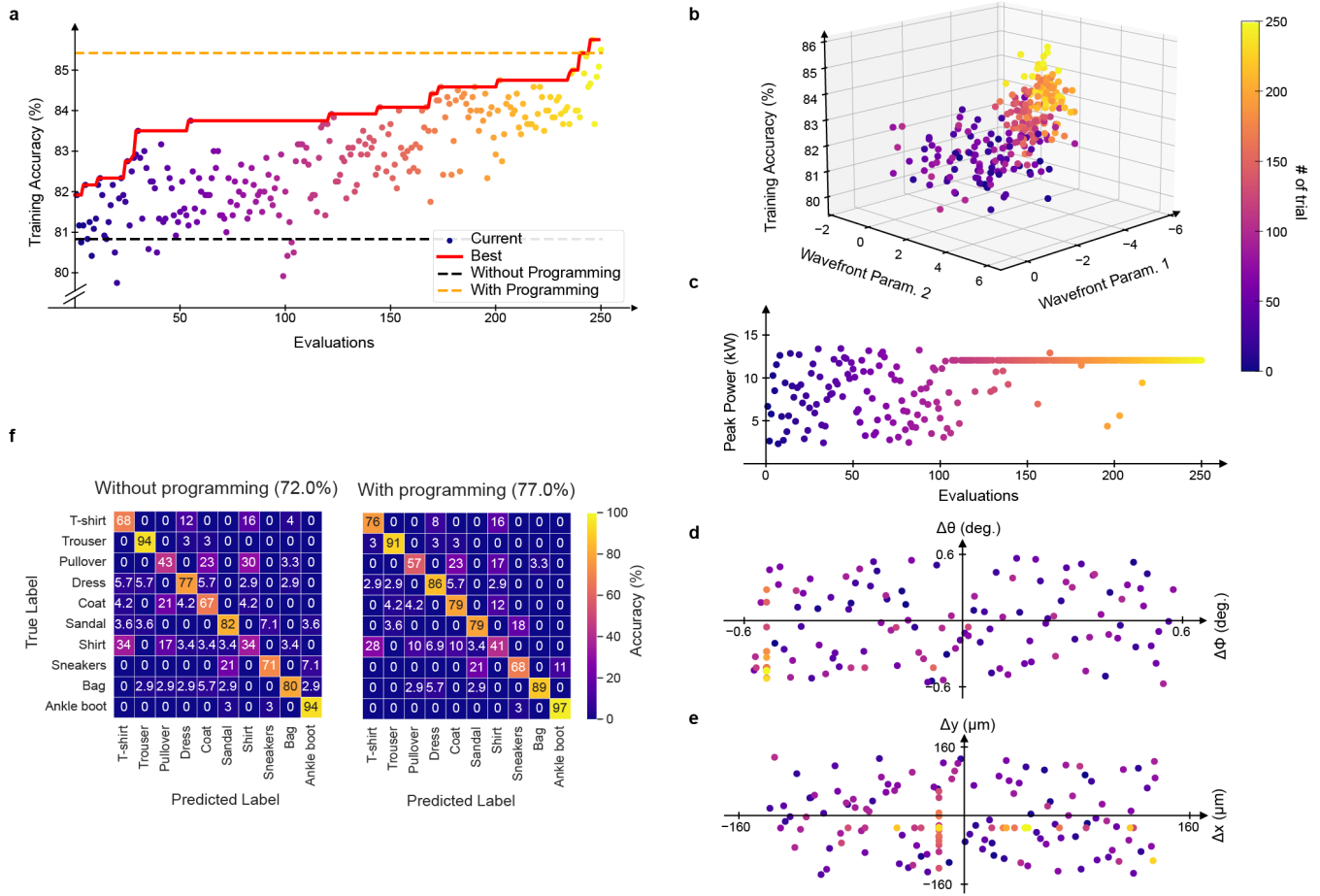

*Figure 2 Programming the MMF propagation for higher classification performance on Fashion-MNIST dataset. a. The training accuracy during the progress of the programming procedure. The horizontal line labeled "Without Programming" shows the accuracy level when PPs are set to zero and "With Programming" indicates the level when the PPs found by the programming algorithm are used. The colors of circles indicate their sequence in the training. b. Relation between wavefront shaping parameters and training accuracy. 46 different wavefront shaping parameters are shown in two dimensions by means of random projection into 2 dimensions for visibility. c. The peak power of pulses during the programming procedure d. The change of the diffraction angle on the SLM, in horizontal and vertical directions. e. The shift of image on the SLM in horizontal and vertical directions f. Confusion matrix and average accuracy on the test set, without and with the programming of the transform*

As shown in Figure 2, for the first 105 iterations, the surrogate optimization algorithm broadly samples the parameter space to create the initial mapping between PPs and the performance metric. After this phase, an area with the potential of yielding the best result is selected and sampled in finer steps. Gradually, the changes become smaller, and the algorithm converges to a solution. Figure 2b provides a closer look at the progression of the process, where each data point represents an iteration, the 2-dimensional random projection of 46 wavefront shaping parameters versus the training accuracy. The initial homogenous sampling of the parameter space and final fine-tuning can be observed. Similarly, Figure 2c shows that after exploring various levels of optical intensity, hence nonlinearity, the convergence led to a higher light intensity for obtaining a more efficient nonlinear optical transform. This is also made possible by converging to a preferred oblique excitation of the fiber as shown in Figure 2d, allowing for stronger coupling to higher order fiber modes as well, hence benefiting from the multimodality of the fiber. Overall, programming the optical propagation by optimized PPs improved the classification accuracy both on the training and the test sets by about 5% and reached 77% accuracy on the test set. This result is on par with the 79.1% accuracy of the 5-layers digital convolutional neural network(CNN) LeNet-5[41], which is trained with the same dataset on a GPU (for details see Supplementary Method). Later we present another approach for programming the propagation, in which PPs are combined with data through convolution, and 79.0% test accuracy is reached on the same task.

## All-Optical Computing with Propagation in Optical Fiber

In Figure 2 the PPs were optimized to modify the optical transform inside the MMF to improve the performance of the combination of the optical system and the digital readout layer. To further demonstrate the programming capacity of our approach, the inference on input samples was done all-optically without any RWs by only using the center location of the output speckles as shown in Figure 3. For a binary classification problem, the input is classified as either "0" or "1" depending on which side of the classification line the center of the output beam resides. Hence, only 3 parameters are required at the output for defining any line in two dimensions. Figure 3 illustrates the programming procedure for all-optical classification of the dataset consisting of 1200 training and 300 test chest radiography images, equally sampled from patients with and without COVID-19 diagnosis[42].

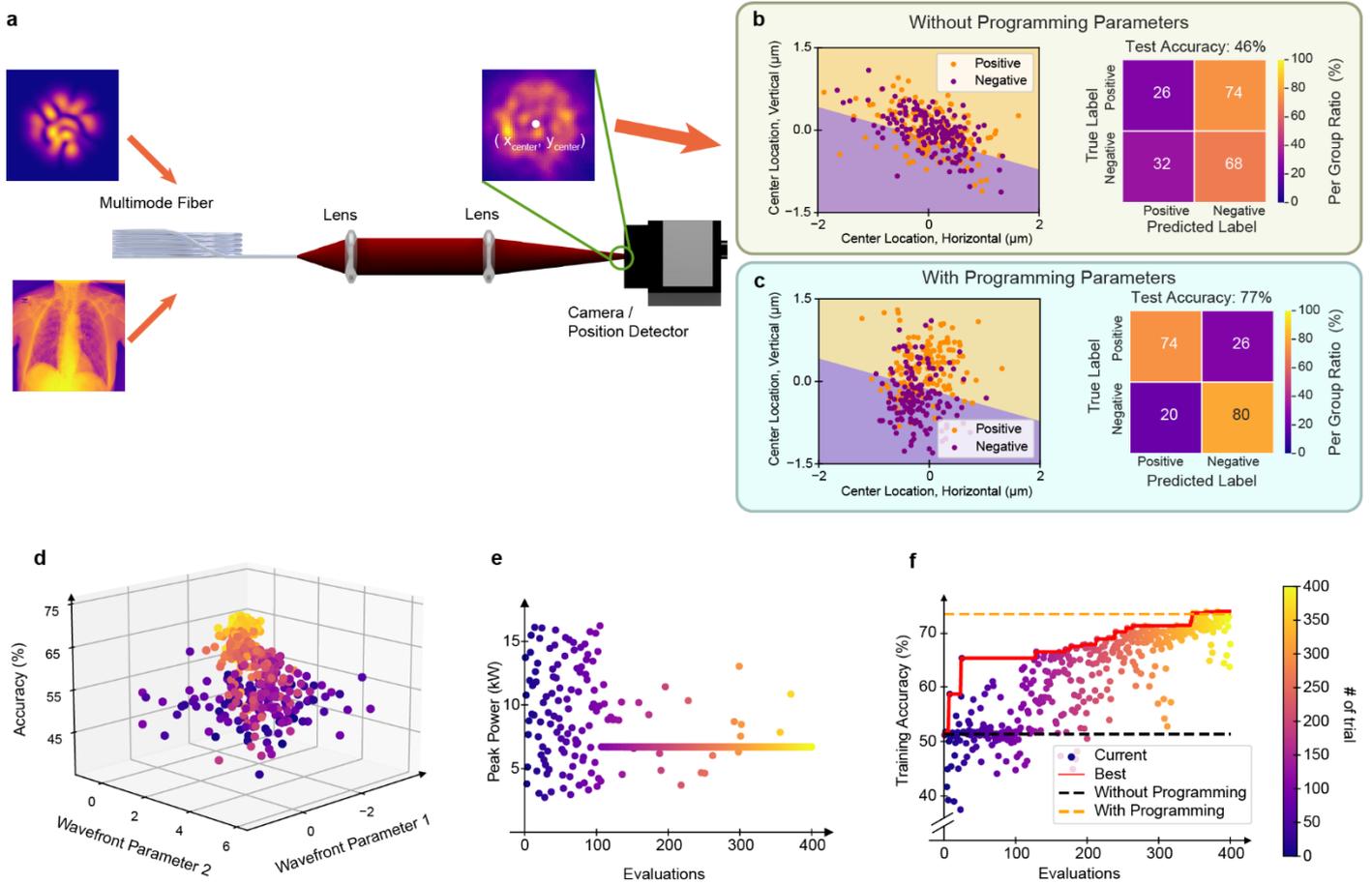

*Figure 3 Programming procedure for all-optical classification of chest radiographs. **a.** The schematic of the experiment, the data, and the control pattern are sent together to the SLM, the fiber output pattern is imaged onto a camera. **b,c.** The distribution of the beam center locations and corresponding confusion matrices for the test set, without and with the programming of the transform. **d.** The distribution of training accuracies with respect to the selection of wavefront shaping parameters. **e.** Selected power levels for each iteration of the programming procedure. **f.** Progression of training accuracy during training. The colormap relates the color of circles to their sequence in the training, it applies to d-e-f.*

Before programming the system, a linear classifier received the distribution of center locations without any control patterns and drew a classification boundary between positive and negative samples. As the transform is random, the classifier could only produce training and test accuracies around 50%. Then, the decision boundary is kept the same and the training accuracy is improved by optimizing the PPs on the SLM with the surrogate model to separate the center location distributions for samples with positive and negative labels. This procedure improved the accuracy on the test set from 46% to 77%, as shown in Table 1. This performance compares favorably to LeNet-5, which uses about 82,000 parameters. This is especially significant considering that the proposed method can be realized all-optically with only 55 TPs. Similarly on a new task of classifying skin lesions of equally sampled benign (nevus) and malignant (melanoma) case images[43], 61.3% test accuracy could be achieved.

Table 1 Comparison between neural networks and all-optical classification system

| Network Structure | Total Number of Parameters | Operations per Sample on Digital Computer (FLOP) | Accuracy on Melanoma dataset (%) | Accuracy on COVID-19 dataset (%) |
|---|---|---|---|---|
| LeNet-5 | 82826 | 1175640 | 64.9 ± 2.0 | 74.6 ± 3.0 |
| MMF + classification with output location (with programming) | 55 | 2029 | 61.3 | 77.0 |

## Different Wavefront Shaping Approaches for Programming the Optical Transform

For the two experiments shown previously, the optical transform was programmed through the multiplication of fields by encoding data and PPs (i.e., $D$ and $WF$ terms in Eqn.2, respectively) as shown in Figure 2 and detailed in Figure 4.a. In this section, we demonstrate that the transform can be achieved with two additional ways of wavefront shaping as depicted in Figure 4.

For the modification of the phase, the control pattern was again formed as described in Eqn.2. Thus, this function is elementwise digitally multiplied with the input image from the dataset and placed on the SLM as visualized in Figure 4.f. For the N-th sample of the dataset, the field diffracted by the SLM becomes $E_{SLM}^N(x,y) = \exp(i(D^N(x,y) * Arg(WF(x,y))))$. After optimization of the PPs, the test accuracy on the subset of Fashion-MNIST reached 78%.

Alternatively, convolutional filters can be used to amplify or attenuate different parts of the angular spectrum of the field, hence its mode decomposition inside the MMF. Importantly, convolutional filters can be applied fully optically by filtering in the Fourier plane[44]. Figure 4.k. depicts that convolution can also program the nonlinear propagation and could reach 79% test accuracy on the same dataset when each element of the convolution kernel is set as PPs. Hence, the $c \times c$ convolution kernel could be written as

$$A = \begin{bmatrix} a_1 & \cdots & a_c \\ \vdots & \ddots & \vdots \\ a_{c^2-c+1} & \cdots & a_{c^2} \end{bmatrix}$$

in terms of programming $c^2$ PPs. Then, the field modulated by the convolution filtered N-th sample of the dataset is $E_{SLM}^N(x,y) = \exp(i(A \circledast D^N(x,y)))$. Similarly on the 1200 training and 300 test samples from the MNIST-digits dataset 94.3% accuracy, which is comparable to the 94.9% accuracy of a 7-layer digital ANN with ~420,000 parameters, could be reached with the same approach demonstrating that different wavefront shaping strategies could realize the enhanced interactions within the optical fiber.

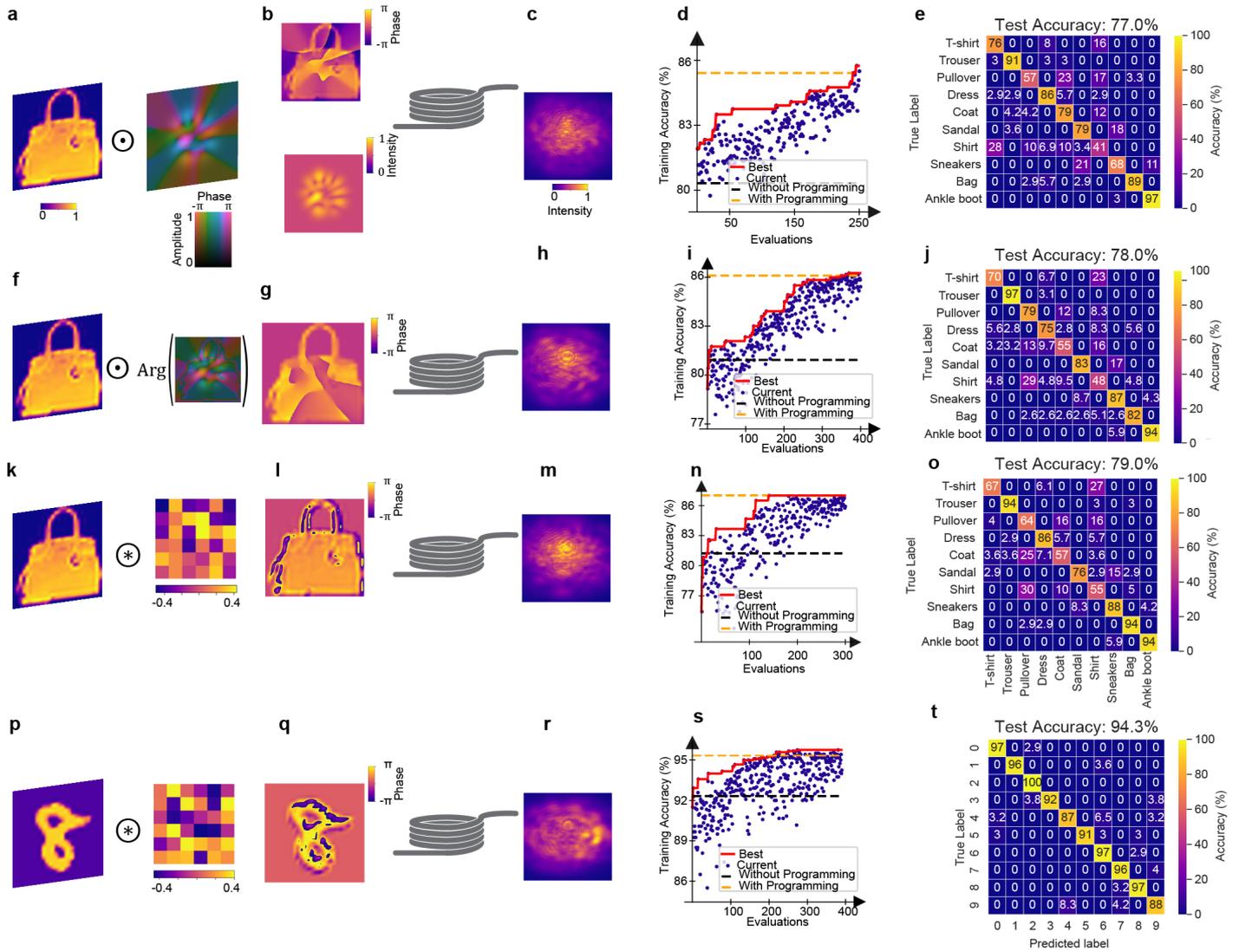

*Figure 4 Programming the optical transform by using phase addition and amplitude modulation(a), multiplication with phase(f), and convolution(k,p). **b,g,l,q** shows an example of programmed patterns on the SLM, for g,l and q the intensity is not modulated. **c,h,m,r** shows recorded intensity patterns after the propagation inside the optical fiber for the given input pattern. **d, i, n, s** depicts the progression of training accuracies during programming iterations. The confusion matrices on **e, j, o, t** illustrate the classification performance of the programmed optical transform with different methods.*

## Transferring Programming Parameters Across Different Tasks and Datasets

The optimization of PPs from scratch requires processing the selected dataset on the experimental system more than 100 times while modifying the PPs. With the approximately 50 frames per second rate, this process takes a few hours for a dataset of 1500 images. In addition to switching to faster optoelectrical devices, the ability to transfer previously optimized PPs to new tasks or datasets would boost the practical utility of this approach as a general-purpose component that can be quickly deployed on different problems. This ability is first demonstrated with the reusability of PPs on different tasks for the same dataset. After finding the set of optimal PPs for the task of classifying the gender of the person in an image from Celebrity Face Attributes Dataset(CelebA)[38], the same set of PPs is used for determining the age of the person. The only training required for the transfer between tasks is the determination of RWs without any new surrogate optimizations involving the optical system. Table 2 compares the performances of the PP transfer between age and gender tasks with fully programming the system for each task separately, showing that the test accuracy with the parameter transfer follows the accuracy of programming from scratch. Without programming the fiber, the test accuracy on the age classification task is 59.0% and 2026 RWs are used. After optimizing an additional 52 PPs (wavefront shaping and experimental parameters, reaching 2078 TPs), the test accuracy reaches 67.0%, performing better than a digital 7-layer CNN with about 412,000 parameters. When these optimized 52 PPs are used on the gender classification task on the same dataset only with re-training of the RWs, the accuracy on the new task is 76.0%, which is similar to the 76.3% achieved by programming the system from scratch with the gender database. The same findings hold true when the initial programming is done on the gender task and parameters are transferred to the age task.

Table 2 The performances of different CNNs and optical computing methods on the CelebA dataset

| Network Structure | Total Number of Parameters | Operations per Sample on Digital Computer (FLOP) | Test Accuracy on Age Task | Test Accuracy on Gender Task |
|---|---|---|---|---|
| LeNet-1 | 914 | 149132 | 61.7±0.2 | 67.3±0.4 |
| LeNet-5 | 82146 | 1174248 | 63.0±0.4 | 75.2±1.6 |
| 7-layer Convolutional NN | 411794 | 65163532 | 65.3±0.1 | 80.1±0.7 |
| MMF + linear output layer | 2026 | 4050 | 59.0 | 69.0 |
| Programmed MMF for Age Task + linear output layer (trained for the corresponding test task) | 2078 | 6075 | 67.0 | 76.0 |
| Programmed MMF for Gender Task + linear output layer (trained for the corresponding test task) | 2078 | 6075 | 64.7 | 76.3 |

Furthermore, we find that the optimized PPs can also be transferred to a different dataset requiring only a short corrective programming. The PPs optimized for the COVID-19 classification dataset with the all-optical approach are transferred to the task of classifying skin lesions between benign (nevus) and malignant (melanoma) case images[43] (Figure 5). However, directly transferring the PPs from former to latter resulted in a test accuracy of 47.67%, which is similar to a random prediction. In corrective programming, a smaller set of parameters (11 in total) are designated for optimizing the previously acquired set of PPs. These 11 parameters are combined with 52 PPs by repetition of each element multiple times and element-wise addition. Thus, an optimal set of PPs is found in the proximity of the initial expectation by optimizing in a lower dimensional search space. Decreasing the dimensionality enables a convergence in fewer iterations. Compared to the complete programming of the system in 300 iterations, corrective programming starts from a similar initial accuracy and after 80 iterations instead of 300, reaches the same final test accuracy.

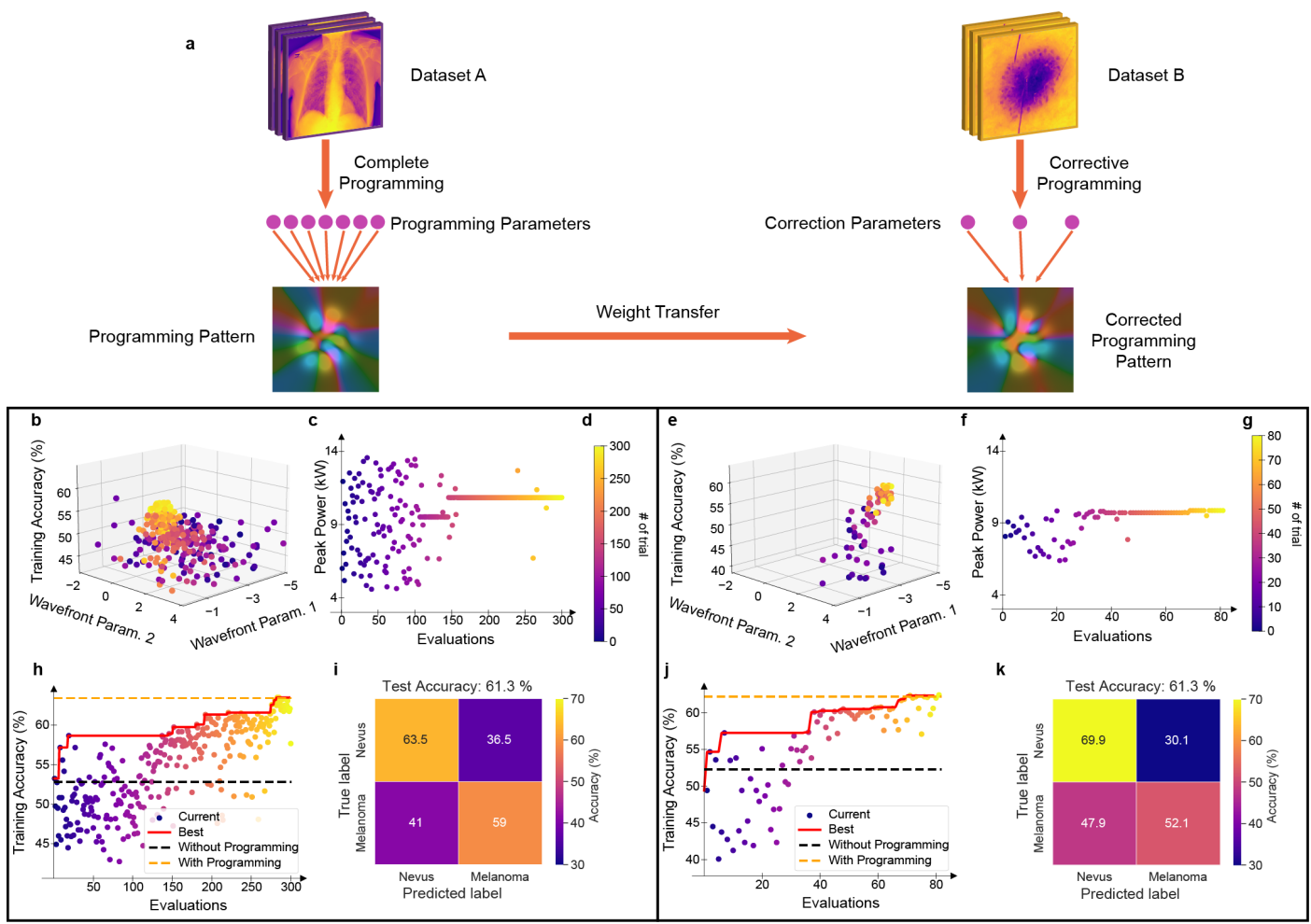

*Figure 5 Using previously dedicated parameters on a new dataset with corrective programming. **a**, The procedure of transferring the PPs. **b,c,d,h,I**, belongs to the experiment when the PPs fully programmed without any prior knowledge. **e,f,g,j,k**, belongs to the corrective programming of parameters. **b,e**, Relation between wavefront shaping parameters projected to 2 dimensions and the training accuracy. **c,f**, The peak power of pulses at the fiber entrance. **d,g**, The color bar for coding the iteration number related to each*

*data point on b,c,h and e,f,j. **h,j,** The training accuracy during the progress of the programming procedure. **i,k,** Confusion matrix and average accuracy on the test set.*

# Discussion

## Computation Speed and Energy

The speed of inferences is limited by the refresh rate of the liquid crystal (LC) SLM. This limitation can be overcome by switching to a faster wavefront shaping method, for instance utilizing commercial digital micromirror devices, which can reach 30000 frames per second[45]. Since the number of modes in the MMF is much smaller than the number of pixels on commercial SLMs, different lines of the SLM could be scanned with the beam by a resonant mirror, allowing up to 25 million samples per second data input rate. Moreover, the fixed complex modulation or convolution operations can be implemented with optical phase masks, bringing the digital operation count further down. Similarly, instead of the digital readout layer, a broadband diffractive element can realize the linear projection step. As it is analyzed in detail in Supplementary Note 4 and visualized in Figure 6, implementing the same optical computer with a selection of commercially available, high-speed equipment such as digital micromirror devices and quadrant photodiodes, 25 TFLOP/s performance could be reached with a total power consumption of 12.6 W, which is significantly lower than 300 W consumption of a GPU with a comparable performance[46].

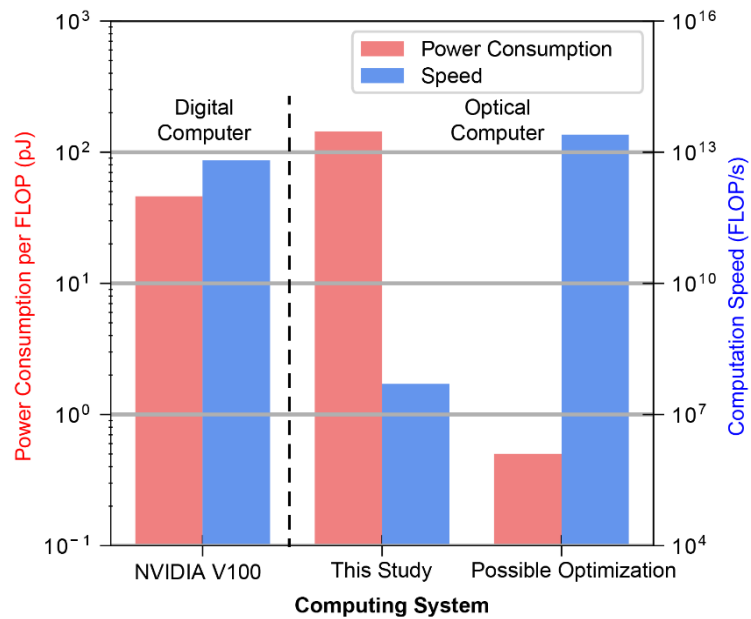

*Figure 6 **The power efficiency and speed comparison between different computational approaches.** The possible optimization refers to incorporating a digital micromirror device, a resonant mirror, and an optical phase mask in the optical computer.*

## Stability and Reproducibility

The reproducibility of experiments is crucial for consistent comparison between different sets of PPs during programming and long-term usability of determined PPs. To investigate reproducibility, the inference experiment is repeated every 5 minutes for the same PPs and RWs on the same task over 15 hours. As shown in Supplementary Figure 1, the first and final test accuracy is the same, and the standard deviation of the test accuracy over time is 0.3%, indicating a very stable experimental inference.

In conclusion, programming nonlinear propagation inside MMFs with wavefront shaping techniques can exploit complex optical interactions for computation purposes and achieve results on par with multi-layer neural networks while decreasing the number of parameters by more than 97% and potentially consuming orders of magnitude less energy. This shows the capacity of nonlinear optics for providing a solution for the exponentially increasing energy cost of the machine learning algorithms. Our work also shows that the performance of a reservoir computer can be improved dramatically through the programming methods we describe in this paper, which can be applied to other modalities as well.

# Methods

## Experimental setup

The experimental system shapes the wavefront of a pulsed IR laser beam, couples it into a graded-index fiber, then collimates and sends it to two different imaging arms. The laser pulses with 10 ps length, 125 kHz repetition rate, and 1030 nm center wavelength are provided by a mode-locked ytterbium fiber laser (Amplitude Laser, Satsuma). The intensity of the laser beam is controlled from the computer via a half-wave plate mounted on a motorized rotation stage (Thorlabs PRM1) followed by a polarizing beam splitter. Then, the beam is incident on the reflective phase-only two-dimensional spatial light modulator with 1920-by-1152 pixels and the pitch size (Meadowlark HSP1920-600-1300). The 8-bit SLM is calibrated to provide phase modulation between 0 to 2π at each pixel location, for transmitted pixel values between 0 to 255. The laser beam has a circular shape with approximately 6.7 mm diameter on the SLM surface, therefore all images shown on the SLM are converted to grayscale, upsampled to 520-by-520 to reach the size of the beam, and has the format of 8-bit unsigned integer. To isolate diffracted beam from the undiffracted portion, all images are shown on the SLM after elementwise addition with a blazed grating phase pattern and its phase depth is varied over the beam area to control the diffraction efficiency for each location, hence achieving amplitude modulation on the diffracted beam. The first-order diffracted beam is clearly separated from the undiffracted beam after propagating 150 mm and focused by a plano-convex lens with a focal length of 15 mm. The input facet of commercially available graded-index MMF (OFS, bend-insensitive OM2) of 50 μm core diameter, 0.20 NA, 5 m long, is placed in the focal plane of the lens and its position is fine-tuned with a 3-dimensional alignment stage. After propagating through the fiber, the beam is collimated with a 20 mm focal length plano-convex lens. Then, it is separated into two beams by a non-polarizing 50-50 beam splitter. The transmitted beam is imaged onto a monochrome CMOS camera (FLIR BFS-U3-31S4M-C) with an achromatic doublet lens of 100 mm focal length. This camera is used for directly inspecting the beam shape and for calculating the center location of the beam. The reflected beam again goes through a beamsplitter, one of the branches arrive at an InGaAs power sensor (Thorlabs S145C), and the other branch is reflected from the diffraction grating (600 lines/mm, Thorlabs GR25-0610) and is incident onto the same model of CMOS camera. Neutral density filters are placed before the cameras in the optical setup to avoid saturation. The detailed schematic of the experiment is shown on Supp. Fig 5. All the electrical equipment is connected to the same general-purpose computer via USB and PCI-E ports.

## Implementation of Programming Parameters

During this study, the programming of propagation is combined with the data on a digital computer, and the combinations were converted to optical signals, even though the combination of PP-controlled shaping and data is also realizable in a fully optical manner by means of a fixed 2-dimensional plate with a pre-determined spatially-varying distribution of phase and magnitude transfer properties. Three ways to program the propagation are presented. In addition to those programming methods based on wavefront shaping, for each experiment, 6 additional parameters controlling light intensity ($I$), displacement of the data on the SLM ($\Delta x$, $\Delta y$), diffraction angle on the SLM with respect to the optical axis of the system ($\Delta\theta$, $\Delta\phi$), and focal length of defocusing ($f$) are optimized.

The first programming method is illustrated in Figure 4.a. and performs an elementwise multiplication of fields due to the data and programming. The programming pattern is formed by the linear combination of analytically calculated propagation modes of the GRIN MMF. For determining the programming pattern as a combination of $N$ modes, $N$ parameter pairs are selected each for real and imaginary coefficients. Therefore the programming complex pattern is formed as $P(x,y) = \sum_{i=1}^{23} F_i(x,y) * (A_i + jB_i)$, where $A_i$ and $B_i$ are the coefficients for the $i$-th mode and $F_i(x,y)$ is the scalar field of the $i$-th mode. These mode fields could be expressed in terms of Laguerre polynomials[47], $L_p(x)$: $F_i(\rho,\phi) =$

$\sqrt{\frac{p!}{\pi(p+|m|)!}} \frac{\rho^{|m|}}{\rho_0^{|m|+1}} e^{-\rho/2\rho_0^2} L_p^{|m|}\left(\frac{\rho^2}{\rho_0^2}\right) e^{im\phi}$. $p$ and $m$ are radial and angular numbers of the modes, $\rho_0^2 = \frac{R}{k\sqrt{2\Delta}}$, $R$ is the radius of the fiber, $k$ is the wavenumber for the center of the fiber, and $\Delta = \frac{n_1^2 - n_0^2}{2n_1^2}$, relative refractive number difference. The input data, $D(x,y)$, is scaled to be between 0 and $2\pi$, then the combination of the programming pattern and other PPs constitutes the diffracted electric field: $E(x,y) = \sqrt{I} \ |P(x - \Delta x, y - \Delta y)| \exp\left[j(D(x - \Delta x, y - \Delta y) + \arg(P(x - \Delta x, y - \Delta y)) + \frac{(x-\Delta x)^2 + (y-\Delta y)^2}{f} + k\cos(\Delta\phi)\sin(\Delta\theta) x + k\sin(\Delta\phi)\sin(\Delta\theta) x\right]$.

In the second case, shown in Figure 4.f, the phase of the programming pattern is pointwise multiplied with the data, in this case, the diffracted electric field is $(x,y) = \sqrt{I} \ \exp\left[j(D(x - \Delta x, y - \Delta y) \arg(P(x - \Delta x, y - \Delta y)) + \frac{(x-\Delta x)^2 + (y-\Delta y)^2}{f} + k\cos(\Delta\phi)\sin(\Delta\theta) x + k\sin(\Delta\phi)\sin(\Delta\theta) x\right]$.

For the convolutional programming (the third case) in Figure 4.k, the $c \times c$ convolution kernel, $C(x,y)$ is formed with $c^2$ PPs. Then, the total field is

$$E(x,y) = \sqrt{I} \exp\left[j(\sum_{x_m}^{c}\sum_{y_m}^{c}(D(x - \Delta x - x_m, y - \Delta y - y_m)C(x_m,y_m)) + \frac{(x-\Delta x)^2+(y-\Delta y)^2}{f} + k\cos(\Delta\phi)\sin(\Delta\theta)x + k\sin(\Delta\phi)\sin(\Delta\theta)x\right].$$

## Programming Procedure

The experiment is considered as a whole with the final digital classifier for the determination of PPs. Therefore, the surrogate model is formed directly for the relationship between the PPs and classification accuracy on the training dataset. This computational framework is realized in Python by using scikit-learn and Python Surrogate Optimization Toolbox(pySOT). In the case of training the digital readout layer for classifying according to the output images, the grating dispersed beam is recorded by the camera and downsampled to 45 x 45 by average pooling. Then these downsampled images are flattened to 2025 features, and the ridge classification algorithm is used for the determination of RWs. For all-optical classification experiments, the output beam shape is imaged onto the camera without grating dispersion and the center of mass, $(x_c, y_c)$, of the beam shape, $G(x,y)$, is calculated by $x_c = \frac{\sum_{x=-\infty}^{\infty} x\,G(x)}{\sum_{x=-\infty}^{\infty} G(x)}$, $y_c = \frac{\sum_{y=-\infty}^{\infty} y\,G(y)}{\sum_{y=-\infty}^{\infty} G(y)}$. This is the same information as the one provided by simple beam location sensors.

The surrogate model based on the radial basis function is initiated by sampling 2M+1 Latin Hypercube points, M being the number of parameters to be optimized. Then the search algorithm employing the DYCORS [48] strategy explores the parameter space for the optimal set of parameters.

*Supplementary Information for*

# Programming Nonlinear Propagation for Efficient Optical Learning Machines

Ilker Oguz[1,2], Jih-Liang Hsieh[1,2], Niyazi Ulas Dinc[1,2], Uğur Teğin[1,2], Mustafa Yildirim[1], Carlo Gigli[2], Christophe Moser[1] and Demetri Psaltis[2]

[1] Laboratory of Applied Photonics Devices, Ecole Polytechnique Fédérale de Lausanne (EPFL), Switzerland

[2] Optics Laboratory, Ecole Polytechnique Fédérale de Lausanne (EPFL), Switzerland

## Contents

# Supplementary Note 1: Physical Explanation of the Transform Dynamics

As described in the main text, spatial modulation of the input beam controls the amplitude and phase of the light to be coupled to the different propagation modes of the optical fiber. During propagation, the phase and amplitude of these modes change due to different linear and nonlinear interactions. Multi-mode generalized nonlinear Schrodinger equation provides a comprehensive analysis of these effects (Supp. Eqn. 1) even for ultrashort pulses[1]. The multidimensionality and nonlinearity of these interactions also provide an understanding of why this physical phenomenon can perform as an efficient computer.

$$\frac{\partial A_p}{\partial z} = \underbrace{i\delta\beta_0^p A_p - \delta\beta_1^p \frac{\partial A_p}{\partial t} - i\frac{\beta_2^p}{2}\frac{\partial^2 A_p}{\partial t^2}}_{\text{Dispersion}} + \underbrace{i\sum_n C_{p,n} A_n}_{\text{Linear mode coupling}} + \underbrace{i\frac{n_2 \omega_0}{A} \sum_{l,m,n} \eta_{p,l,m,n} A_l A_m A_n^*}_{\text{Nonlinear mode coupling}} \qquad \text{Supp. Eqn.1}$$

Supp. Eqn. 1 provides a simplified explanation of multimode interactions occurring in the proposed system. Here $A_p$ is the complex coefficient of the p-th normalized mode, where $\beta_k^p$ is the k-th order propagation constant for this mode. $C_{p,n}$ is the linear coupling coefficient between mode p and n, which becomes nonzero when there are non-idealities in the waveguiding structure such as ellipticity, bending, or impurities. $n_2$ is the nonlinear refractive index of the material, $\omega_0$ is the center angular frequency and A is the area of fiber core. Since the propagation length is limited to 5 meters, the higher-order dispersion effects are not shown. Similarly, in the experiment, the Raman scattering is not observed dominantly, hence in the nonlinear mode coupling part, only the part related to the Kerr effect is shown. There, $\eta_{p,l,m,n}$ is the nonlinear coupling coefficient between modes, it mainly depends on the similarity between spatial shapes of modes, and can be calculated as follows[1,2]:

$$\eta_{p,l,m,n} = \int dxdy F_p^* F_l F_m F_n^* \qquad \text{Supp. Eqn.2}$$

Considering the varying strength of interactions between different sets of modes, controlling their coupling explains how implicitly the proposed programming scheme can program the propagation for better performance of the task. In other words, the PPs modify the mapping between input images and mode distributions, where each set of modes has fixed nonlinear coupling weights. This way, PPs select an optimal combination of high-dimensional weights stored already in the medium. Therefore, with a much smaller number of PPs the performance of digital neural networks could be reached.

# Supplementary Note 2: The Stability of the Experiment over Time

Due to the fluctuation in environmental conditions and other noise factors, fluctuation in the laser output power is expected. To decouple this fluctuation from the optical computing setup, light intensity at the output of the multimode fiber is tracked with a power meter and corrections to the angle of polarization before the polarizing beamsplitter are applied according to this reading. This way the light intensity inside the optical fiber can be kept stable. Moreover, when wavefront shaping is applied, the same reading is also used for making sure the intensity of light coupled inside the MMF is directly controlled by the intensity programming parameter. To investigate reproducibility, the inference experiment is redone for the same PPs and RWs on the same task over 15 hours continuously by sending the samples to the optical system. During those experiments, the inference accuracy, the stabilized power level, as well as the average correlation between samples are recorded. This correlation is calculated by the average of correlation coefficients for each input sample, between the initial output beam shape and the output beam shape for the current experiment. As shown in Supplementary Figure 1, the fluctuation in the correlation values and power levels resemble each other closely, and they are both much smaller than 1% over 14 hours. The average beam correlation in 14 hours is 99.9% with a 0.04% standard deviation and compared to its maximum, the power level is 99.7% on average with a 0.14% standard deviation. Consequently, the accuracy fluctuations are minimal, with the first and final test accuracy being exactly the same at 83.7%, the average value is 83.8% and the standard deviation is 0.35%. Moreover, single-step readout weight training allows lightweight recalibration, which can further decrease fluctuations by applying once in many experiments.

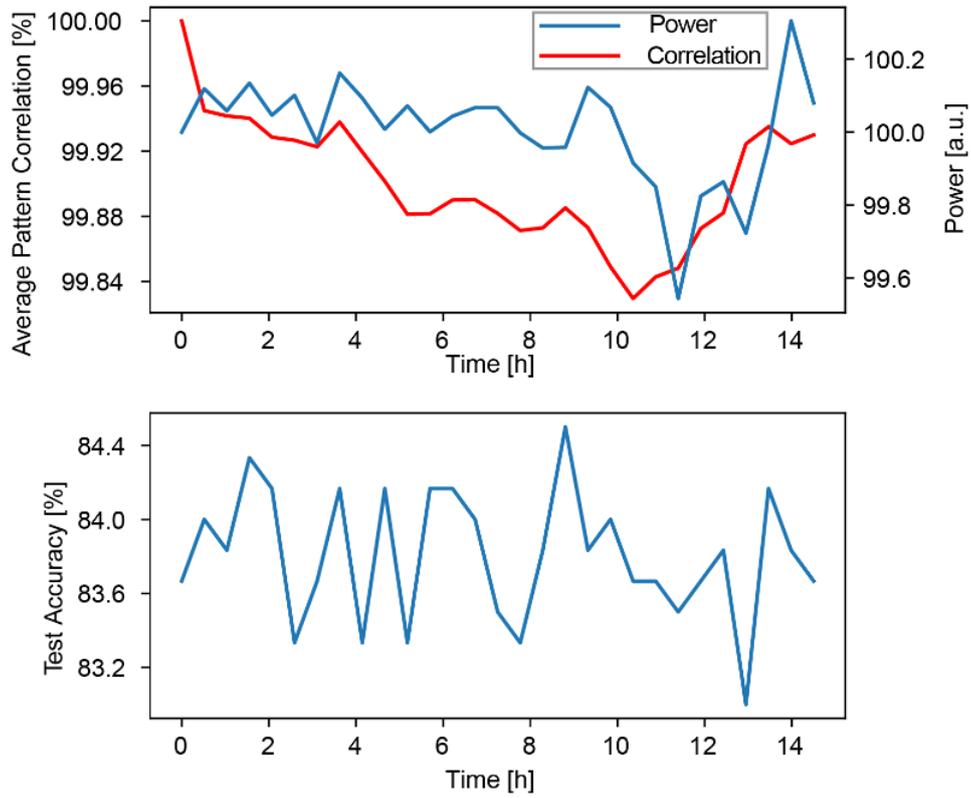

*Supplementary Figure 1 Coupled power, output pattern correlation, and classification accuracy of the system over time*

# Supplementary Note 3: Dependence of the Performance on Pulse Length and Power Level

As the main determining factor for the strength of the nonlinear effects, the peak power also modifies the nature of the data transform through the optical fiber. As it was previously shown[2], there is an optimal peak power level for a given task, while there is not enough nonlinearity below this level, above this level beam cleaning effects start to become prominent and hinder the efficiency of data processing.

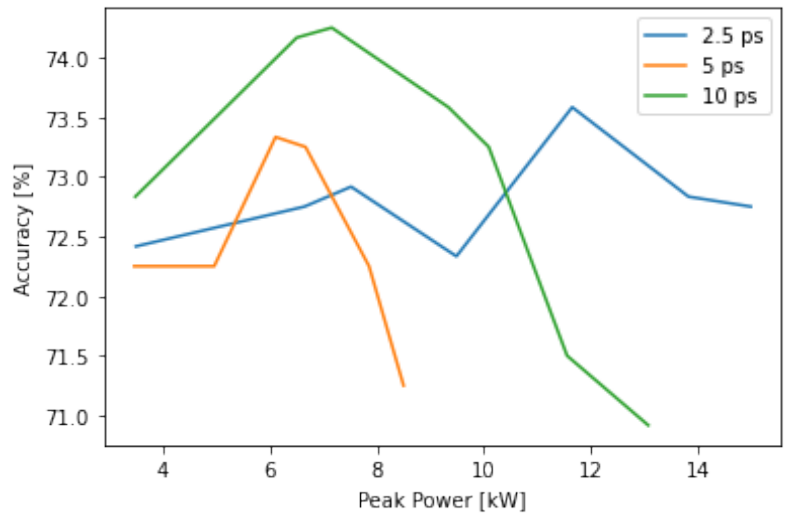

*Supplementary Figure 2 The dependence of training accuracy on pulse peak power and pulse length on the CelebA dataset, the age classification task*

Our findings in Supplementary Figure 2, without programming the propagation inside MMF, corroborates those findings, showing that for different pulse lengths, the optimal level of peak power and corresponding training classifications stay about the same. Moreover, in this case, the length of the fiber is not scaled with the pulse length, so the strength of dispersion effects on those pulses are different. This difference can be one of the reasons for the small discrepancy observed between different experiments. In addition, the output beam shapes shown on Supplementary Figure 3, which corresponds to the same set of experiments with Supplementary Figure 2, indicates that the dynamics of transform stay similar at different pulse lengths as long as the peak power is also scaled.

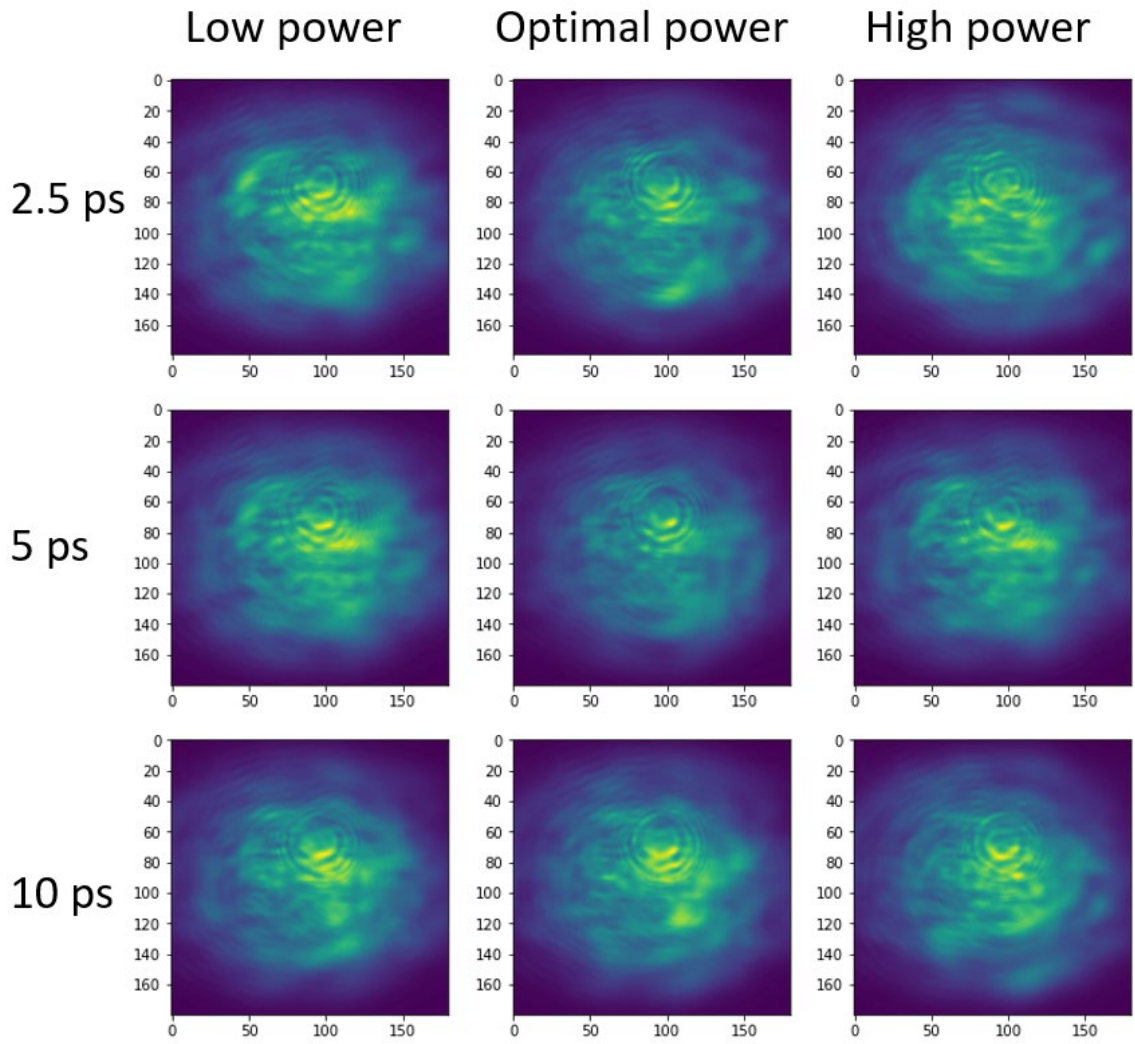

*Supplementary Figure 3 The output beam shapes for the same input image for different power and pulse length levels*

# Supplementary Note 4: Operation Speed and Power Consumption Analysis of the Computation Method

## Dimensions and Operation Count

The experimental results on different datasets indicate that programming MMF propagation can achieve similar accuracies with digital NNs requiring ~1 MFLOP/sample. Therefore, for the rest of the document for each image processed with the optical system, this process will be accounted as equivalent to 1 MFLOP of operation on the GPU. And for each operation, the input is assumed to be 22x22 pixels with 8 bit-depth and the output is 45x45 pixels with 8 bit-depth. These resolutions are determined by the supported modes on the fiber. On the input side, the diffraction-limited resolution is sampled by 2 input pixels and on the output side, the sampling is about 4 pixels per diffraction-limited point size because of the spatially displayed temporal information thanks to the diffraction grating.

## Optical Power Consumption

For optimal performance, the optical system requires 10 ps long pulses with about 10 kW peak optical power inside the fiber. Therefore with 100 nJ(=$10^{-11}$x$10^4$nJ) per pulse, the optical power efficiency for the equivalent digital computation is 0.1 pJ/FLOP.

## Power Consumption for Optoelectrical Conversion

Currently, the data is encoded to optical pulses via a liquid crystal phase modulator controlled by an analog high-speed circuitry, it has 1920 x 1152 pixels with 8-bit precision, each pixel sized 9.2 μm x 9.2 μm working at ~ 50 Hz or 20 ms per frame. The SLM consumes in total (24V,1A) 24 W continuously. Therefore, each refresh consumes around 0.5 J (=24Wx20ms), or 217nJ/pixel (0.5/(1920x1152)). For optimal utilization of the SLM, in which all pixel information couples to a fiber, by using multiple fibers or only activating the subgroup of pixels with light incident on them, the consumption could be brought to 22x22 pixels x217 nJ/pixel = 105 μJ/image.

We can replace the phase-only LC SLM with a DMD, for instance, Texas Instruments DLP9500, which can show 23148 patterns per second with a 1920x1080 resolution and 4.5 W electrical consumption[4]. Therefore, the power consumption per frame is 196 μJ and per mirror is 94 pJ. For modulation of 22x22x8 bits per input image, the energy cost of modulation would be 0.4 μJ.

For recording output beam shapes, a commercial CMOS camera is used, the camera has a resolution of 720 x 540 pixels and can reach 522 frames per second, consuming 3W. This amounts to 5.7 mJ for each full frame recording and 15 nJ per each pixel readout, or for 45x45 pixel output images, 30 μJ /image. Again, this camera can be replaced with another commercially available, but more power-efficient version; for instance, LUX 1310, can read out with 1.4nJ/pixel, which can decrease the consumption down to 2.8μJ/image.

The energy consumption of laser is calculated by taking fiber coupling, SLM diffraction, and laser efficiency into consideration. The efficiency of coupling into MMF is around 50%, for both the LC SLM and the DMD light diffraction efficiency greater than 70% and a Yb-doped fiber-based femtosecond laser can convert the electrical energy to light pulses with about 3.3% efficiency[5]. Therefore, the electrical power cost of a 100 nJ pulse inside the fiber is calculated to be 8.7 µJ.

*Supplementary Table 1 The energy budget breakdown of the proposed optical computing method*

| Optoelectronic Device | Energy Consumption per Image-Consumption of All Pixels on the Device | Energy Consumption per Image-Consumption Scaled to Required Pixels |
|---|---|---|
| LC SLM | 0.5 J | 105 µJ |
| DMD | 196 µJ | 0.4 µJ |
| Camera | 5.7 mJ | 30 µJ |
| Laser Consumption | 8.7 µJ | 8.7 µJ |
| Total Consumption-LC SLM | 506 mJ | 144 µJ |
| Total Consumption-DMD | 5.9 mJ | 39 µJ |

The energy consumption of each component of the experiment is provided on Supplementary Table 1. In the two columns of the table, two different ways of accounting for the energy are presented, the first column accounts for the case where consumption due to all of the possible pixels is included. In the second column, the energy consumption of the device is scaled to the proportion which is actually used. Moreover, in the calculation of total consumption two alternatives are presented, the one with the LC SLM is the current experimental setting, while the DMD option is the case when LC SLM is switched with a DMD while keeping the experiment exactly the same. In the latter, by turning on only the necessary pixels per image energy consumption could be lowered to 39 µJ per image, corresponding to 39 pJ/FLOP. In comparison, benchmark results show that an NVIDIA V100 GPU provides inferences with ResNet-50 neural network at an energy consumption rate of 46 pJ/FLOP[6].

Switching from an LC SLM to a DMD could improve the inference speed by nearly 3 orders of magnitude, reaching up to 23000 frames/second, and allowing 23GFLOP/s. However, as it can be observed in the optical telecommunication systems, the transfer and manipulation of information in the optical domain supports conveniently GHz-level rates. To demonstrate that this potential indeed encompasses the proposed computation method, in the following section we describe an implementation that can reach the processing speed of state-of-the-art GPUs, with a favorable efficiency.

# High-Speed Low-Power Consumption System

The energy efficiency and high-performance computing potential of the proposed method can be realized with a high-speed system as shown in Supplementary Figure 4.

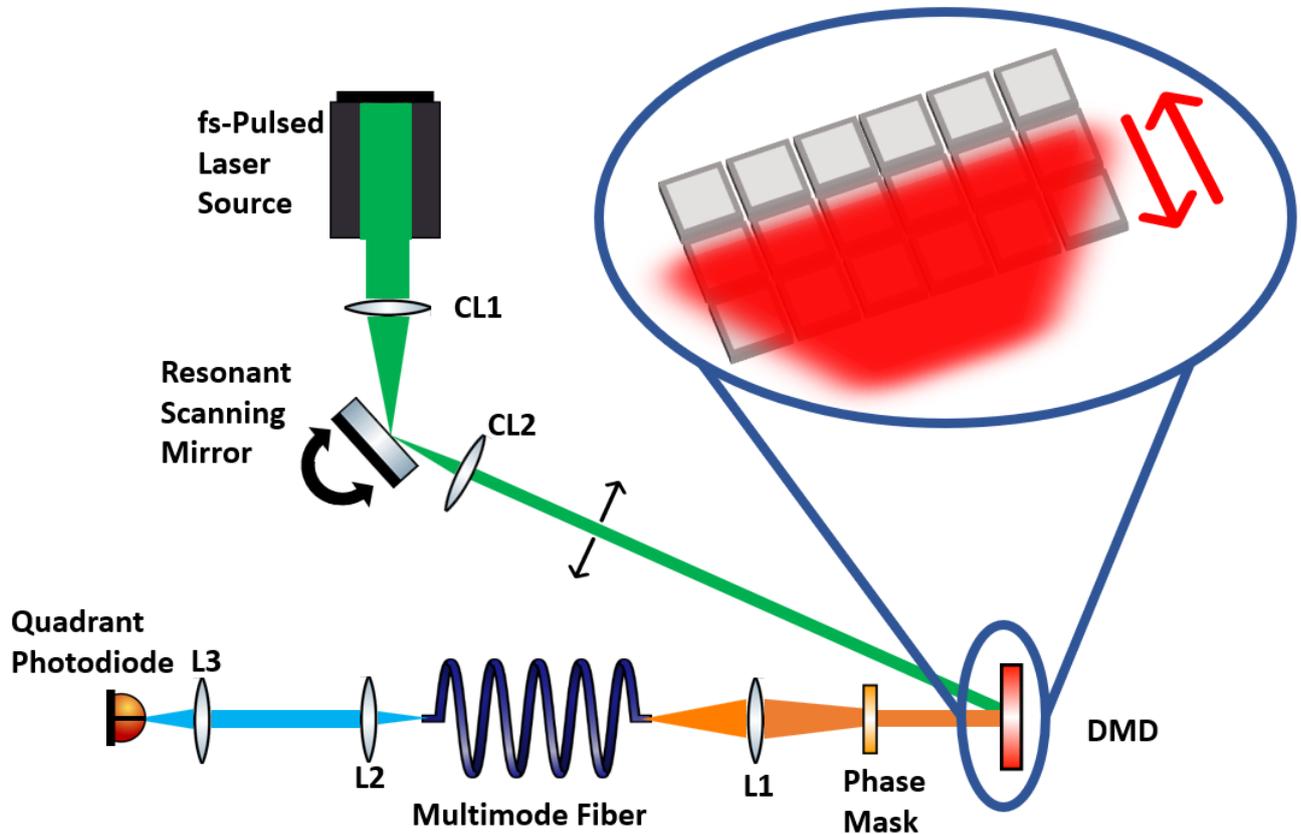

*Supplementary Figure 4 The experimental schematic of a high-speed implementation of the proposed method*

For reaching high data processing speed efficiently, some of the system parameters and components should be changed without changing the main principle of computation. These changes are:

1- Pulse length
2- Spatial modulation device
3- Fiber length
4- Readout device.

As the strength of optical nonlinearities is proportional to the peak power of the optical pulse, the required average optical power for the same strength of the nonlinearity could be decreased by working with a shorter pulse. Moreover, with the shorter pulse, the nonlinear interactions happen at a shorter length scale and the fiber length could be scaled proportionately, creating a more compact setup. In the high-speed system, instead of the current 10 ps pulses, 300 fs can be used. Then the fiber length can be scaled down to 17 cm instead of 5 m. This can decrease the light source related consumption by 33 times, down to 264 nJ/image.

In our current proof-of-principle system a liquid-crystal SLM is utilized. Even though their ability to directly control the phase makes them ideal prototyping tools, the slow response of liquid crystals (a few milliseconds) limits the modulation speed with the LC SLMs. On the other hand, digital micromirror devices offer a much faster alternative with their millions of fast switchable mirrors (few microseconds). To benefit fully from DMD's speed and the high number of pixels, while only a few hundreds of channels exist in the multimode fiber, we propose using a narrow beam and sweeping the lines of the DMD with a resonant scanner as shown in Supplementary Figure 4. For instance, the utilization of a ~11.5 kHz resonant mirror (Novanta Photonics, CRS 12) whose upwards and downwards cycles are synchronized to 23148 Hz rate refreshing with a DMD (TI DLP9500), could achieve the processing of individual lines on the DMD, each line consisting of 1920 pixels would encode an input to be processed by the system. Therefore, 1920/240 = 8 bits of information can be coupled to each one of the 240 channels of the fiber and 23148x1080 ≈25 million samples per second can be shown. For the conversion of 1-dimensional data and programming to 2-dimensions, a scattering phase mask can be used[7]. After coupling into the fiber and propagation, the output beam location, which was shown to be programmable to directly provide inference results without any need for a digital readout layer, can be tracked with a sectioned photodiode such as Excelitas C30665GH-4. Therefore, 25 million inferences per second could be performed, and equivalently 25 TFLOP/s processing speed could be reached. In comparison, during the benchmarking on the ResNet50 neural network, as NVIDIA V100 GPU has a peak consumption of 300 W[6], with 46 pJ/FLOP, it can provide 6.5 TFLOP/s.

*Supplementary Table 2 The energy budget breakdown of the high-speed implementation of the optical computing method*

| Optoelectronic Device | Power Consumption (W) |
|---|---|
| DMD | 4.4 |
| Resonant Mirror | 1.5 |
| Quadrant Photodiode | 0.1 |
| Femtosecond Laser | 6.6 |
| Total Consumption | 12.6 |

In addition to the high-speed computation, the proposed implementation is power efficient. As the breakdown can be seen in Supplementary Table 2, while performing inferences with 25 million samples per second and the power consumption would be 12.6 W, corresponding to 0.5 pJ/FLOP efficiency.

In summary, the overall energy consumption of the system depends on the energy cost of input modulation and reading out of pixels, in addition to the energy required to create a pulsed laser beam. As the pulse repetition rates of femtosecond lasers can reach up to GHz levels, the computation speed is effectively limited by the slowest of the optoelectronic conversion devices. The proof-of-principle system with a selection of optoelectronic equipment not being optimized for power consumption can achieve computing with power consumption similar to the GPU. Replacing only the SLM of this setup with a DMD is expected to improve further the energy efficiency to an advantageous level and speed, since DMD is faster than LC SLMs and per bit modulation cost is lower. Scanning different locations of the DMD with the laser beam and reading out the inference result all-optically is expected to bring down the energy consumption 2 orders of magnitude while providing significantly faster computation.

The potential of this computing system can be further explored by decreasing the dominant power expenditure on optoelectronic modulation by employing silicon photonics based modulation, which can achieve an efficiency of 0.1 pJ/bit at 10 Gbit/s [8]. In addition to further improving the per bit efficiency 3 orders of magnitude compared to the DMD, with its high speed, it removes the need for the resonant mirror. In this case, the dominant power expenditure would be from the femtosecond laser and this expenditure could be again decreased with an optimal fiber, with which nonlinearities could be enhanced more than 10 times for the same peak power level or could provide the same amount of nonlinearities with 10 times lower peak powers[9].

Furthermore, for the proposed improvements, fiber mode number, hence the input and output resolutions are assumed to be kept the same. However, as we discuss in the next section, the efficiency of the system can be increased by scaling the number of modes.

## Discussion on Scalability of the Approach

In the previous sections, different implementations are proposed to improve the speed and power efficiency while the fiber dimensions are kept the same. Another possible approach would be to increase the number of propagation modes, $N = \frac{\alpha}{\alpha+2} n_0^2 k^2 R^2 \Delta$ [1]. This number scales with the square of the core radius, $R^2$, with the square of the numerical aperture of the fiber ($NA^2 = 2\Delta n_0^2$), and the inverse square of the light wavelength ($\lambda = 2\pi/k$). As N could be increased with simply a different selection of the fiber and light source, the number of effective calculations is expected to scale with at least $N^2$ and potentially even steeper. This expectation stems from the interaction terms in the Multimode Generalized Nonlinear Schrodinger Equation (Supp. Eqn. 1). Even in the absence of nonlinear interactions, linear interactions create some amount of coupling from each mode to every other mode for each infinitesimal propagation step. With nonlinear interactions, these interactions become products of four mode coefficients, which create a number of interactions on the $N^4$ order. On the other hand, the sample input rate and energy consumption are proportional to N, for instance, if modulation with M features takes t time, with the proposed laser scanning method, modulation with 2M features for 2N modes will take 2t. Similarly, for the same level of nonlinearity the optical peak power per mode (P/N) should be kept the same, so for a 2N mode fiber, the optical power consumption should be 2P, while the number of operations should scale to at least 4C from C. Then, increasing the dimensionality as much as possible should clearly improve the power efficiency and operation speed. Even with the same wavelength, a 100-fold increase in the dimensionality and an improvement in the operation speed and power efficiency at the same ratio can easily be achieved by switching the current fiber with the core diameter of 50 μm to another commercially available fiber with a 500 μm core diameter.

## Supplementary Method: Comparison with Digital Neural Networks

To create a one-to-one comparison between the presented optical computation method and electronics-based digital implementations, we used the same portions of the same dataset for training and testing. The digital neural networks were implemented with the Keras software library. In addition to LeNet-1 and LeNet-5 [3] implementations, a larger version of LeNet-5 with 7 layers and higher numbers of filters per layer was benchmarked. This neural network has the same input size and nonlinear activations as the LeNet-5 while it consists of 2 consecutive 2-dimensional convolution layers with 48, 3x3 kernels, a 2-dimensional Max Pooling, 2 consecutive 2-dimensional convolution layers with 96, 3x3 kernels, a 2-dimensional Max Pooling, a 2-dimensional convolution layers with 192, 3x3 kernels, a 2-dimensional Max Pooling, a dense layer with 512 neurons, and finally a dense layer with neurons as many as output classes.

For each one of the models and datasets, the training is done with 5 different random starting of the neural network, the test set accuracies are recorded, and the mean and the standard deviation of these trials are reported.

## Supplementary Figure: Detailed Experimental Setup

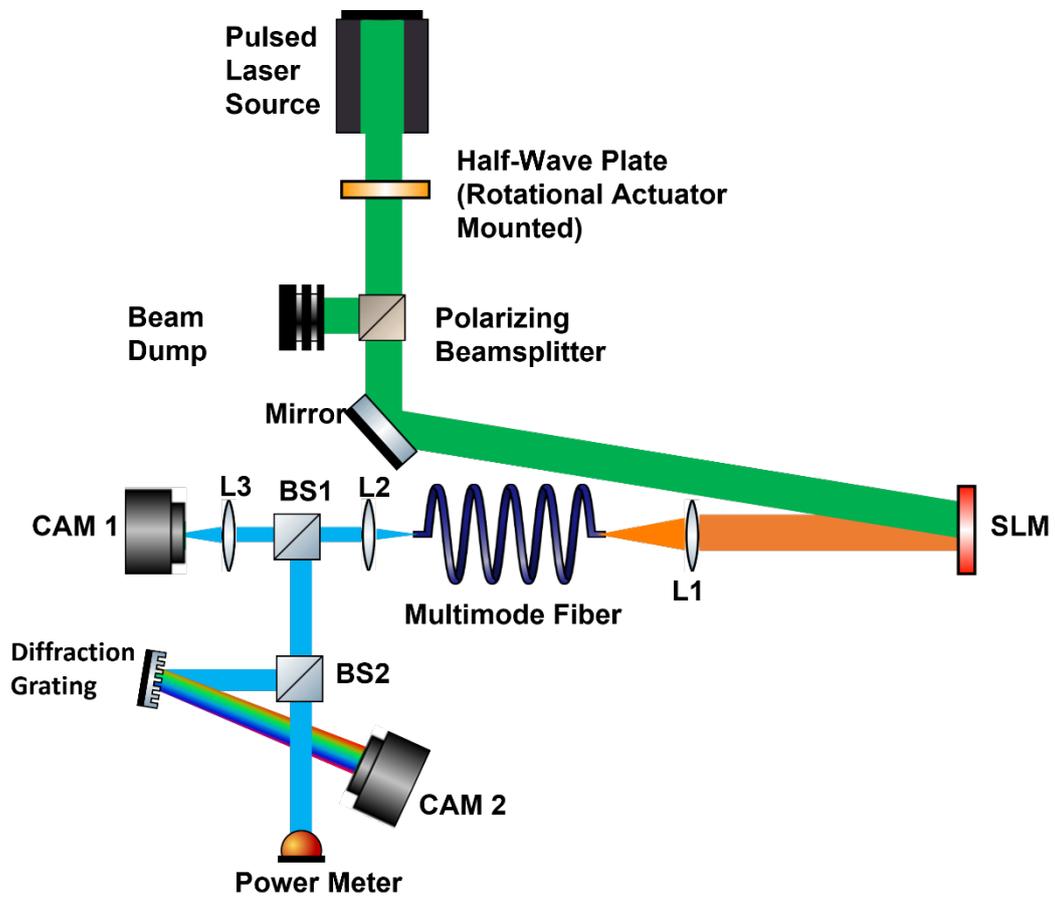

*Supplementary Figure 5 The detailed schematic of the experimental setup*

# Supplementary References